\documentclass[amsmath,amssymb,aps,prl,twocolumn,showkeys,superscriptaddress,10pt,floatfix,showpacs]{revtex4-1}

\usepackage{tabularx}
\usepackage{makecell}
\usepackage[dvipsnames]{xcolor}
\usepackage{fdsymbol}
\usepackage{graphicx}
\usepackage{bm}
\usepackage{siunitx}
\usepackage{hyperref}
\usepackage{chemformula}

\usepackage{xcolor}
\hypersetup{
    colorlinks,
    linkcolor={red!50!black},
    citecolor={blue!50!black},
    urlcolor={blue!80!black}
}

\begin{document}
\title{Pressure-induced suppression of ferromagnetism in CePd$_2$P$_2$}
\author{T.\ A.\ Elmslie}
\affiliation{Department of Physics, University of Florida, Gainesville, FL 32611, USA}
\author{D.\ VanGennep}
\affiliation{Department of Physics, University of Florida, Gainesville, FL 32611, USA}
\author{W.\ Bi}
\affiliation{Advanced Photon Source, Argonne National Laboratory, Argonne, IL 60439, USA}
\affiliation{Department of Geology, University of Illinois at Urbana-Champaign, Urbana, IL 61801, USA}
\affiliation{Department of Physics, University of Alabama at Birmingham, Birmingham, AL, 35294, USA}
\author{Y.\ Lai}
\affiliation{National High Magnetic Field Laboratory, Florida State University, Tallahassee, Florida 32310, USA}
\author{S.~T.\ Weir}
\affiliation{Physics Division, Lawrence Livermore National Laboratory, Livermore, CA 94550, USA}
\author{Y.\ K.\ Vohra}
\affiliation{Department of Physics, University of Alabama at Birmingham, Birmingham, AL, 35294, USA}
\author{R.\ E.\ Baumbach}
\affiliation{National High Magnetic Field Laboratory, Florida State University, Tallahassee, Florida 32310, USA}
\author{J.\ J.\ Hamlin}
\email{Corresponding author: jhamlin@ufl.edu}
\affiliation{Department of Physics, University of Florida, Gainesville, FL 32611, USA}

\begin{abstract}
The correlated electron material CePd$_2$P$_2$ crystallizes in the ThCr$_2$Si$_2$ structure and orders ferromagnetically at \SI{29}{K}.
Lai \textit{et al.}~\cite{lai_ferromagnetic_2018} found evidence for a ferromagnetic quantum critical point induced by chemical compression via substitution of Ni for Pd.
However, disorder effects due to the chemical substitution interfere with a simple analysis of the possible critical behavior.
In the present work, we examine the temperature - pressure - magnetic field phase diagram of single crystalline CePd$_2$P$_2$ to \SI{25}{GPa} using a combination of resistivity, magnetic susceptibility, and x-ray diffraction measurements.
We find that the ferromagnetism appears to be destroyed near \SI{12}{GPa}, without any change in the crystal structure.
\end{abstract}

\maketitle

\section{Introduction}
When a ferromagnetic transition is suppressed by a clean control parameter such as pressure, typically, the second-order phase transition changes to first-order at a critical value of the control parameter~\cite{pfleiderer_magnetic_1997} and the transition abruptly drops towards \SI{0}{K}~\cite{brando_metallic_2016,kirkpatrick_universal_2012,belitz_first_1999,belitz_tricritical_2005,kirkpatrick_exponent_2015}.
As the system approaches the critical point in a second order phase transition, fluctuations in the order parameter extend to larger and larger length scales, while the order parameter varies smoothly between the ordered and disordered phases.
However, in a first order phase transition, this correlation length does not diverge, and the order parameter changes discontinuously~\cite{belitz_first_1999,janoschek_fluctuation-induced_2013}.
In certain compounds, such as  UGe$_2$ and  ZrZn$_2$, the shift from a second- to first-order transition is accompanied by the appearance of metamagnetic ``wings'' in the phase diagram, in which the ordered phase reappears when a magnetic field is applied~\cite{kaluarachchi_quantum_2018,uhlarz_quantum_2004,kimura_haas--van_2004}.
As pressure increases, the metamagnetic transition is smoothly pushed to higher fields and lower temperatures until it can terminate at a quantum wing critical point (QWCP) at \SI{0}{K}~\cite{belitz_first_1999,belitz_quantum_2017,kaluarachchi_tricritical_2017}.
More complicated scenarios are also possible, where both ferromagnetic and antiferromagnetic or modulated phases are present~\cite{belitz_quantum_2017} as observed in LaCrGe$_3$~\cite{kaluarachchi_tricritical_2017,taufour_ferromagnetic_2018} and CeTiGe$_3$~\cite{kaluarachchi_quantum_2018}.  The complex phase diagrams of such materials represent a critical test of our understanding of quantum phase transitions.

The possibility of these sorts of features make CePd$_2$P$_2$ an interesting compound for study.
The crystal structure of CePd$_2$P$_2$ was first reported in Ref.~\cite{jeitschko_ternary_1983}. 
No further characterization was performed until the work of Shang \textit{et al.}~\cite{shang_tunable_2014}, which reported resistivity, DC magnetization and DC magnetic susceptibility of polycrystalline CePd$_2$As$_{2-x}$P$_x$ for different levels of substitution, and demonstrated a ferromagnetic transition in CePd$_2$P$_2$ at approximately \SI{29}{K}.  
In the same year, Tran and Bukowski~\cite{tran_ferromagnetism_2014} reported on DC magnetic susceptibility, magnetization, specific heat, resistivity, and magnetoresistance measurements on polycrystalline CePd$_2$P$_2$,  and Tran \textit{et al.}~\cite{tran_magnetic_2014} reported AC susceptibility and DC magnetization, also for polycrystalline CePd$_2$P$_2$.
These papers also confirmed ferromagnetic order in CePd$_2$P$_2$ below about \SI{28.4}{K}.
Neutron diffraction and DC magnetization measurements were performed by Ikeda \textit{et al.}~\cite{ikeda_characterization_2015} on a polycrystalline sample, including a magnetically aligned polycrystalline sample, which revealed the magnetic anisotropy of CePd$_2$P$_2$, with the $c$-axis as the magnetic easy axis.
This is confirmed by single crystal work~\cite{drachuck_magnetization_2016}.
The compound CeNi$_2$P$_2$ shares the ThCr$_2$Si$_2$ crystal structure with CePd$_2$P$_2$, but exhibits a non-magnetic ground state~\cite{nambudripad_investigation_1986}, suggesting Pd-to-Ni substitution can drive a transition from magnetic to non-magnetic.
Lai \textit{et al.}~\cite{lai_ferromagnetic_2018} examined this possibility by substituting Ni to replace Pd, revealing a possible ferromagnetic quantum critical point in the temperature-concentration phase diagram.
Here, according to Belitz-Kirkpatrick-Vojta (BKV) theory~\cite{belitz_nonanalytic_1997,belitz_first_1999}, the quenched disorder inherent to the chemical substitution allows the transition to be driven continuously to zero.

Because Ni is isoelectronic with Pd and smaller in size, one can think of Pd $\rightarrow$ Ni substitution as inducing chemical pressure.
Lai \textit{et al.}~\cite{lai_ferromagnetic_2018} compared this chemical pressure effect to the effects of applied mechanical pressure on CePd$_2$P$_2$ up to about 2 GPa.
A small suppression of the Curie temperature with applied pressure was observed indicating a critical pressure for full suppression of the transition well beyond the maximum pressure of that experiment.
Therefore, we undertook to explore the phase diagram of this material at substantially higher pressures.
In particular, we wished to look for the point at which the magnetic transition became first order, as well as any signs of a modulated magnetic phase or metamagnetic wings.
We find that magnetic order appears to be destroyed near 12 GPa.
We find no evidence of a modulated magnetic phase or metamagnetic wings in resistivity measurements between 12 and 20 GPa in fields up to 9 T.

\section{Methods}
Single crystals of CePd$_2$P$_2$ were grown in a molten metal flux according to the process outlined in Ref.~\cite{lai_ferromagnetic_2018}.  
The single crystals had a tendency to break into flat platelets perpendicular to the $c$-axis and were therefore easily aligned with the $c$-axis parallel to applied field.

Alternating current (AC) susceptibility measurements were performed in a Quantum Design PPMS using an Almax-EasyLab Chicago Diamond Anvil Cell (ChicagoDAC).
The magnetic coil system and measurement electronics have been described elsewhere~\cite{vangennep_pressure-induced_2017}.  
In AC susceptibility measurements,  diamonds with \SI{0.8}{\milli \meter} culets were used, and Berylco-25 gaskets were pre-indented to \SI{70}{\micro \meter} from a starting thickness of \SI{260}{\micro \meter}.
The gasket hole diameter was approximately \SI{260}{\micro \meter}, and these gaskets were sealed in a quartz tube under argon atmosphere and hardened in a furnace at \SI{315}{\celsius}.
A solution of 50\% n-pentane to 50\% isoamyl alcohol was used as a pressure transmitting medium~\cite{klotz_hydrostatic_2009}.

When analyzing the AC susceptibility data, a background subtraction was performed for each run, since the signal from the sample is much smaller than the background.  
Background subtraction was performed by subtracting one run from another, always choosing runs which had distant values of $\rm{T_C}$.  
The value of $\rm{T_C}$ was defined as the inflection point of the curve, as determined from second derivative data.

For the resistivity measurements, samples with typical dimensions of $\sim$\SI{70}{\micro \meter} $\times$ \SI{70}{\micro \meter} $\times$ \SI{10}{\micro \meter} were cut from larger crystals and loaded into either an Almax-EasyLab OmniDAC or the ChicagoDAC mentioned above. 
While the ChicagoDAC was used in a Quantum Design PPMS, the OmniDAC measurements were carried out in a custom-made continuous flow cryostat built by Oxford Instruments.
One of the diamonds used was a so-called designer diamond anvil, which is composed of eight symmetrically arranged tungsten microprobes that are encapsulated in high purity homoepitaxial diamond~\cite{weir_epitaxial_2000}.
The designer diamond anvil had a culet diameter of \SI{180}{\micro \meter}, while the opposing anvil had a culet diameter of \SI{500}{\micro \meter}.
Gaskets were made of 316 SS and were preindented to an initial thickness of $\sim$\SI{30}{\micro \meter}.
Quasihydrostatic, soft, solid steatite was used as a pressure medium.
Resistance was measured in the crystalline $ab$-plane using the Van der Pauw geometry with currents of \SI{1}{\milli \ampere}.  
Electrical resistivity measurements performed in the OmniDAC are referred to as ``run A,'' and ChicagoDAC resistivity measurements are designated ``run B.''
In both the ChicagoDAC and OmniDAC, pressure was determined \textit{in-situ} via ruby flourescence~\cite{chijioke_ruby_2005} so that the reported pressures were measured at the corresponding temperatures.

X-ray diffraction measurements were carried out in beamline 16 ID-B of the Advanced Photon Source at Argonne National Lab using a beam with dimensions of approximately \SI{15}{\micro \meter} $\times$ \SI{15}{\micro \meter} and wavelength \SI{0.4066}{\angstrom}.  
Samples were powdered in a mortar and pestle before being loaded into a Symmetric Diamond Anvil Cell (Symmetric DAC) alongside a ruby fragment and small piece of Pt foil for pressure determination.
The Pt equation of state of Holmes \textit{et al}.~\cite{holmes_equation_1989} was used for pressure calibration.
The diamonds had a culet diameter of \SI{500}{\micro \meter}.  
The gasket was made from 316 SS and was preindented to an initial thickness of about \SI{60}{\micro \meter}.  
The gasket hole diameter was approximately \SI{200}{\micro \meter}, and was filled with a pressure medium of 1:1 n-pentane isoamyl alcohol before the sample, ruby and Pt were loaded.  
The cell was pressurized \textit{in situ} via a computer-controlled pressure membrane.

\section{Experimental Results}
\begin{figure}
\includegraphics[width=\columnwidth]{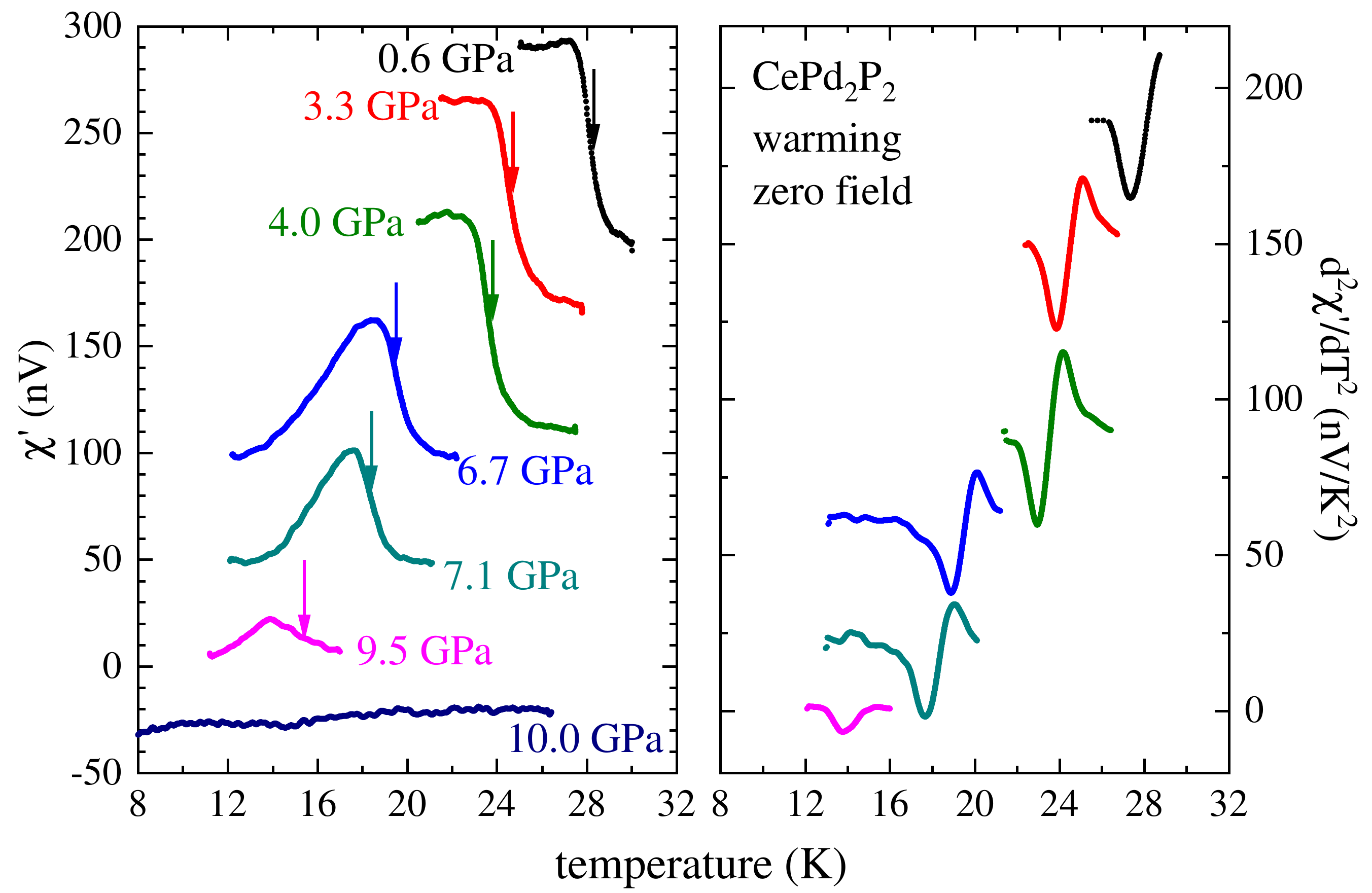}
    \caption{The left panel shows the PM-FM transition of CePd$_2$P$_2$ in the real part of the AC magnetic susceptibility at various pressures.  Note that as pressure increases, the transition temperature decreases, as does the size of the transition.  At \SI{10.0}{\giga \pascal} and above, the transition becomes undetectable in $\chi$.  The transition temperature was defined as the inflection point, as determined by the second derivative of the real part of the magnetic susceptibility, and is indicated by arrows on each curve.  The right panel shows the second derivative of the real part of the AC magnetic susceptibility for each transition.  In both plots, data are offset for clarity.}
 \label{fig:ac}
\end{figure}

Figure~\ref{fig:ac} shows a plot of the real part of the AC magnetic susceptibility of CePd$_2$P$_2$ vs temperature in the vicinity of the transition.
Increasing pressure causes the transition to be suppressed to lower temperatures - from \SI{28.3}{K} at \SI{0.6}{GPa}, to \SI{15.4}{K} at \SI{9.5}{GPa}.
As the transition is suppressed, the magnitude of the anomaly at $\rm{T_C}$ is also reduced.
At \SI{9.5}{GPa} the anomaly is still barely visible, but by \SI{10}{GPa} it is no longer detectable.
The arrows in the figure indicate the criterion used to determine $\rm{T_C}$ and are based on the inflection point, as determined from second derivative data.

The decreasing size of the susceptibility anomaly is consistent with the Rhodes-Wolfarth picture~\cite{rhodes_effective_1963} of itinerant ferromagnetism, which suggests that a decrease in $\rm{T_C}$ will correspond to a decrease in the ordered moment.
There does appear to be a subtle change in the shape of the anomaly in $\chi$ vs $T$.
However, the necessity of background subtraction limits the temperature range of the data that can be compared between the high and low pressure data, which makes it difficult to disentangle changes in the background from changes in the shape of the anomaly.
While it is possible that the nature of the magnetic order changes with pressure, the resistivity measurements (presented below) also do not allow us to make a definite conclusion regarding the possibility of a change in the nature of the magnetic order.

\begin{figure}
\includegraphics[width=0.7\columnwidth]{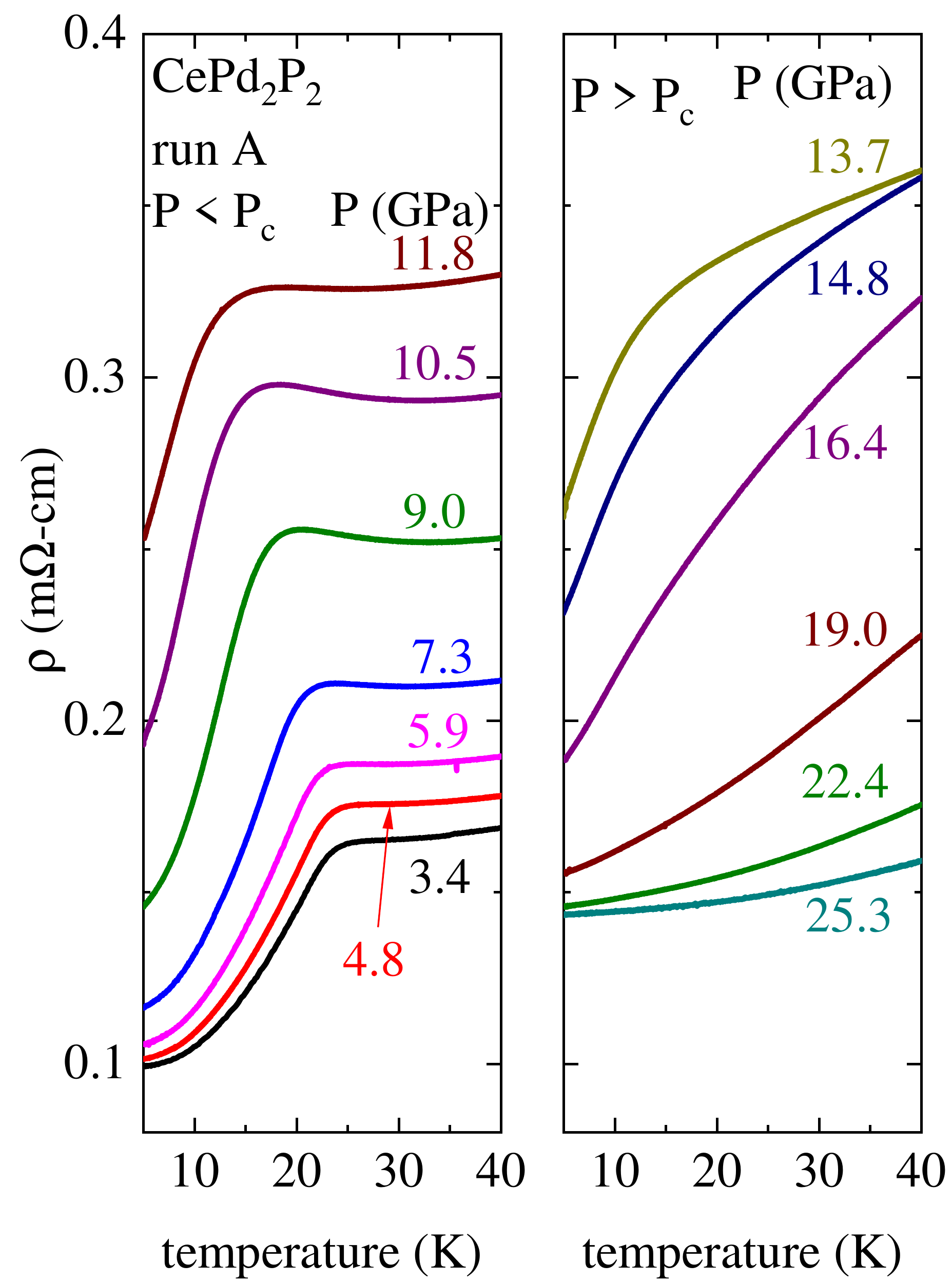}
    \caption{Resistivity as a function of temperature for CePd$_2$P$_2$ run A at pressures below \SI{12}{\giga \pascal} (left) and above (right).  Above \SI{12}{\giga \pascal}, the transition temperature can not be accurately determined.}
 \label{fig:RvsT1}
\end{figure}

\begin{figure}
\includegraphics[width=0.7\columnwidth]{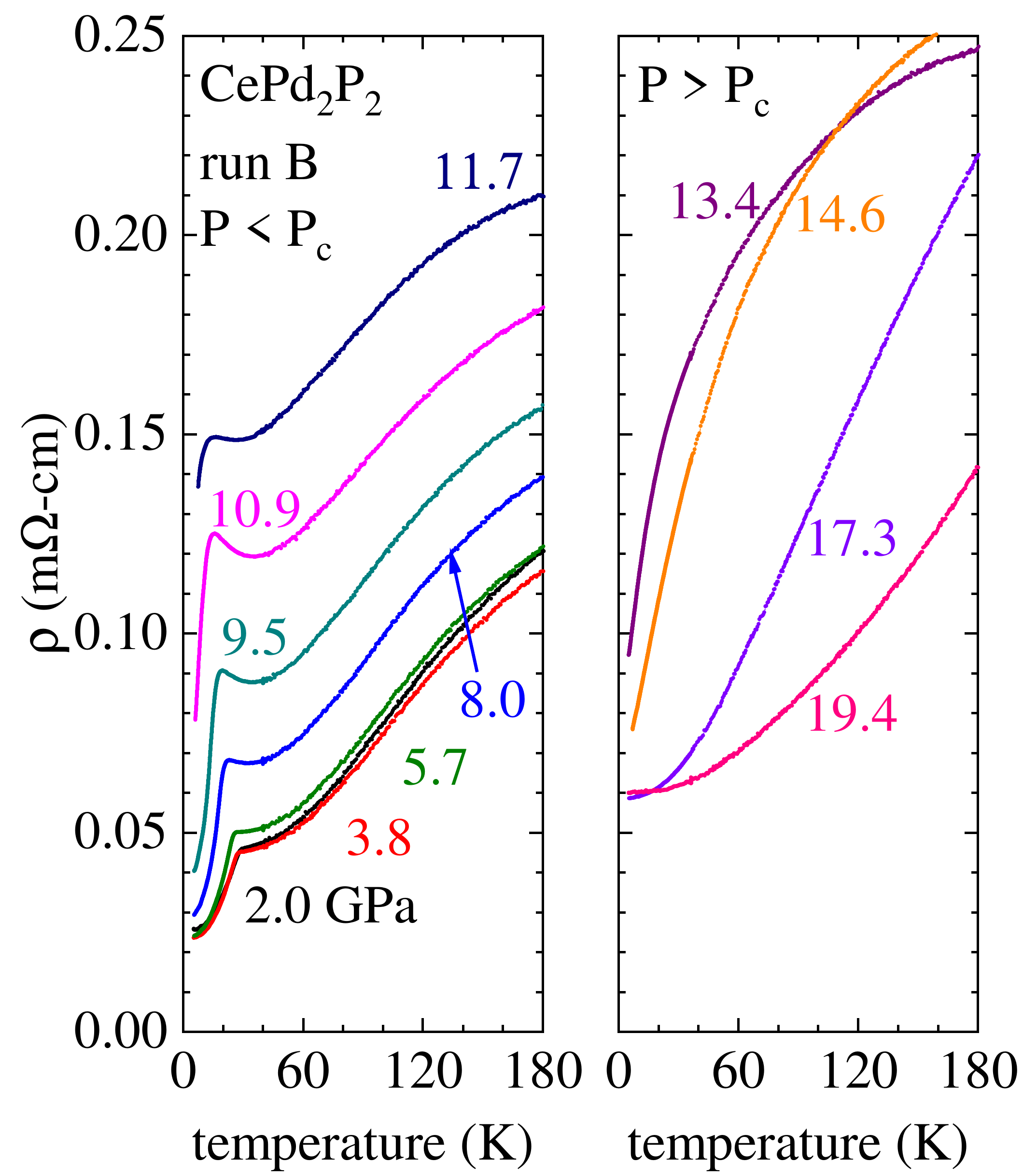}
    \caption{Resistivity as a function of temperature for CePd$_2$P$_2$ run B at pressures below \SI{12}{\giga \pascal} (left) and above (right), showing resistivity behaviour up to \SI{180}{\kelvin}.  }
 \label{fig:RvsT2}
\end{figure}

Figures \ref{fig:RvsT1} and \ref{fig:RvsT2} present electrical resistivity versus temperature data for two different experimental runs, referred to as run A and run B.
In run A, resistivity measurements were performed between \SI{5}{\kelvin} and \SI{40}{\kelvin}, while run B collected data between \SI{5}{\kelvin} and \SI{180}{\kelvin}, demonstrating resistivity behaviour to higher temperature.
The left panels show resistivity curves below \SI{12}{\giga \pascal}, where the transition is easily discernible, and decreases in temperature as pressure increases.  
The plots on the right show data above \SI{12}{\giga \pascal}, where $\rm{T_C}$ can no longer be unambiguously determined.  
In the right-most plot of Fig.~\ref{fig:RvsT1}, the transition is initially still visible (\SI{13.7}{GPa}), but the broadening prevents accurate and reliable determination of $\rm{T_C}$.  
As pressure increases further, the transition disappears entirely and the curvature changes from negative to positive.
We attempted to fit the low temperature electrical resistivity with a power law of the form $\rho = \rho _0 + A T^n$, but found that the data could not be described in this way over a significant range of temperatures.

\begin{figure}
\includegraphics[width=0.9\columnwidth]{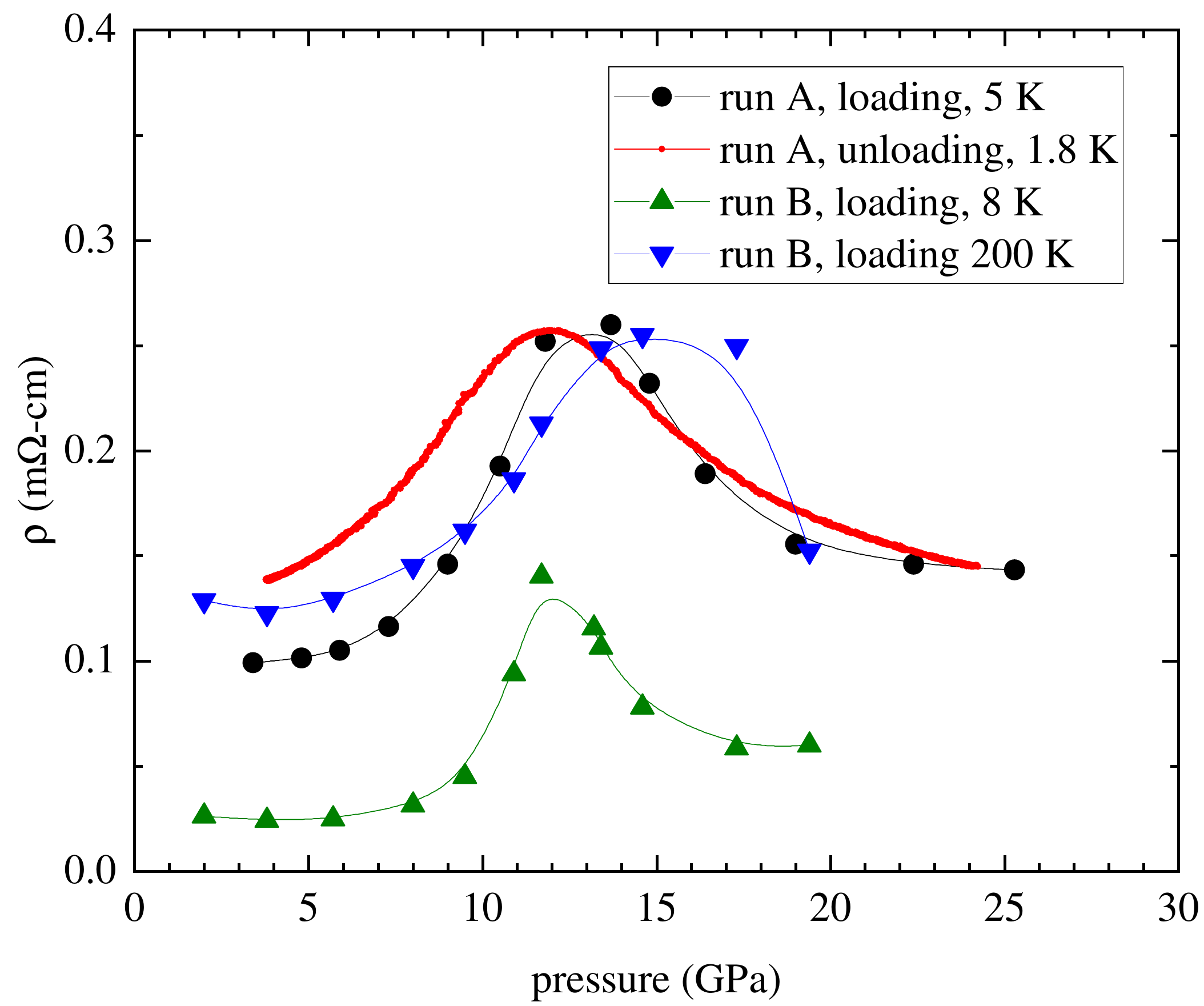}
    \caption{Resistivity as a function of pressure for CePd$_2$P$_2$.  Note the resistivity peak near \SI{12}{\giga \pascal}, close to the pressure where the transition disappears in $\chi$ and $\rho$.}
 \label{fig:RvsP}
\end{figure}

The low temperature resistivity versus pressure curve shows a peak near \SI{12.5}{\giga \pascal}, as demonstrated by the data in Fig.~\ref{fig:RvsP}. 
Data points in the loading curves are synthesized from resistivity versus temperature sweeps at each pressure, while the unloading curve was taken while continuously sweeping pressure at constant temperature.  
The location of the peak displays a hysteresis of about \SI{1}{\giga \pascal} between loading and unloading, and shifts to higher pressures at higher temperatures (at \SI{200}{K} the peak occurs near \SI{15}{GPa}).
The peak appears to coincide roughly with the pressure where the magnetic transition becomes undetectable in $\chi$ and $\rho$.

\begin{figure}
\includegraphics[width=\columnwidth]{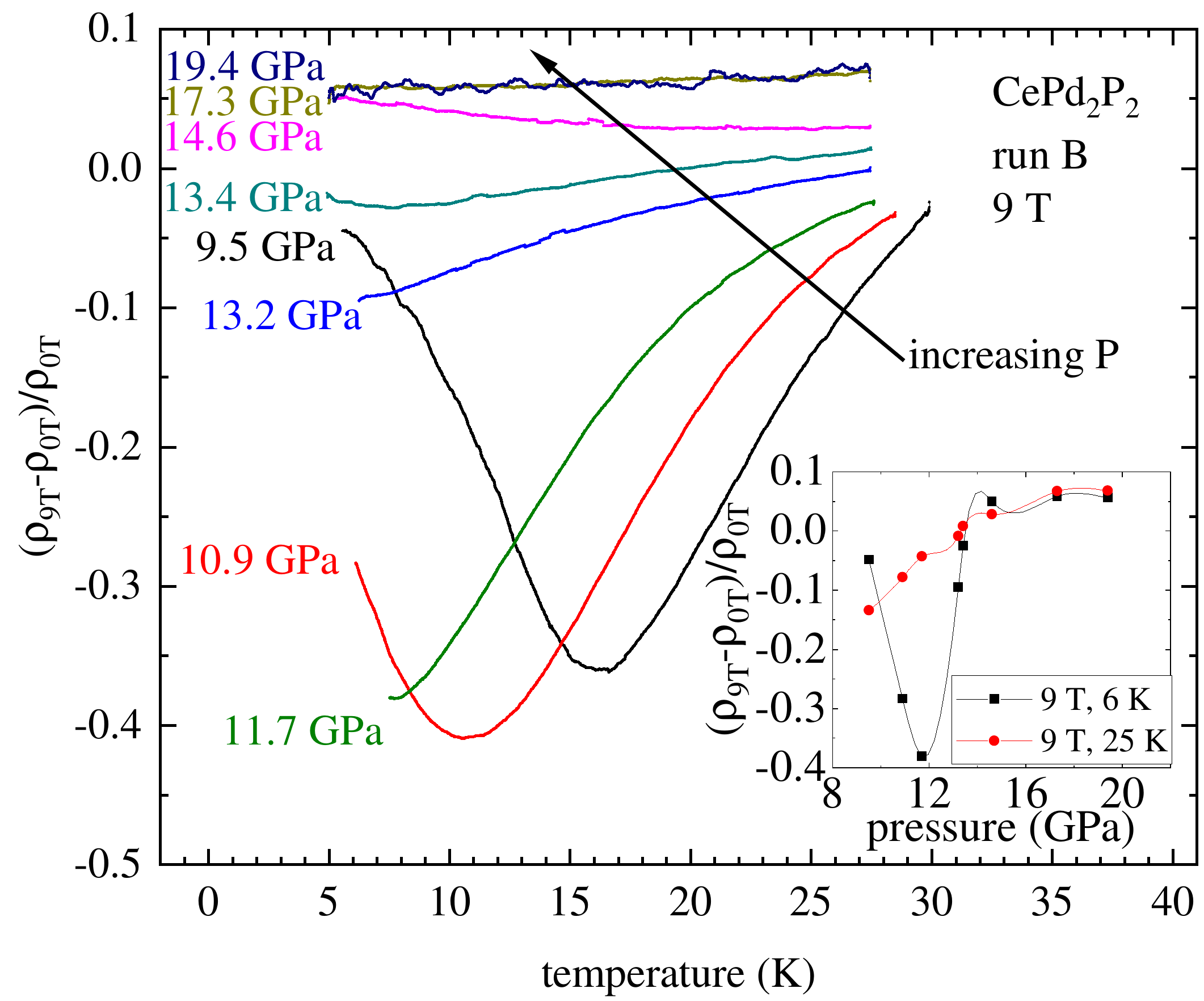}
    \caption{Magnetoresistance as a function of temperature for CePd$_2$P$_2$ at varying pressures.  At low pressure, magnetoresistance is negative and possesses a clear valley near the transition temperature for a given pressure.  Above about \SI{12}{\giga \pascal}, however, this feature vanishes, and as pressure increases further, magnetoresistance shifts from negative to positive above about \SI{13.4}{\giga \pascal}.  The inset shows magnetoresistance of CePd$_2$P$_2$ as a function of pressure at 6 K and 25 K.  At 25 K, above $\rm{T_C}$, magnetoresistance starts negative and gradually increases, becoming positive at higher pressures.  The \SI{6}{\kelvin} data shows a deep minimum at \SI{11.7}{\giga \pascal}, near the critical pressure where the magnetic order appears to vanish.}
 \label{fig:MRvsT}
\end{figure}

Figure \ref{fig:MRvsT} presents the magnetoresistance behavior of CePd$_2$P$_2$ as a function of temperature.
Below \SI{12}{GPa}, as temperature decreases, magnetoresistance becomes large and negative near $\rm{T_C}$.
The temperature at which the magnetoresistance obtains the largest magnitude decreases with increasing pressure, which is consistent with a suppression of the transition to lower temperatures.
As pressure increases above \SI{12}{GPa}, the temperature dependence of the magnetoresistance becomes increasingly flat.
The minimum follows similar behavior to $\rm{T_C}$, disappearing near the critical pressure of \SI{12}{GPa}.
This behavior, in which the magnetoresistance extremum follows the transition temperature, is explained by Yamada and Takada~\cite{yamada_negative_1972} as resulting from fluctuations of localized spins.
The inset of Figure~\ref{fig:MRvsT} presents the magnetoresistance as a function of pressure at \SI{9}{T} for temperatures of \SI{6}{K} and \SI{25}{K}.
The \SI{6}{K} magnetoresistance reaches a minimum at \SI{11.7}{GPa}, close to the value of critical pressure $\rm{P_c}$ as determined by zero field resistivity.
At higher pressures, the magnetoresistance then rises to be nearly identical to the \SI{25}{K} curve.
The \SI{25}{\kelvin} magnetoresistance starts negative and asymptotically approaches zero as pressure increases.
This behavior is consistent with a supression of the magnetic transition near \SI{12}{GPa}.

\begin{figure}
\includegraphics[width=0.8\columnwidth]{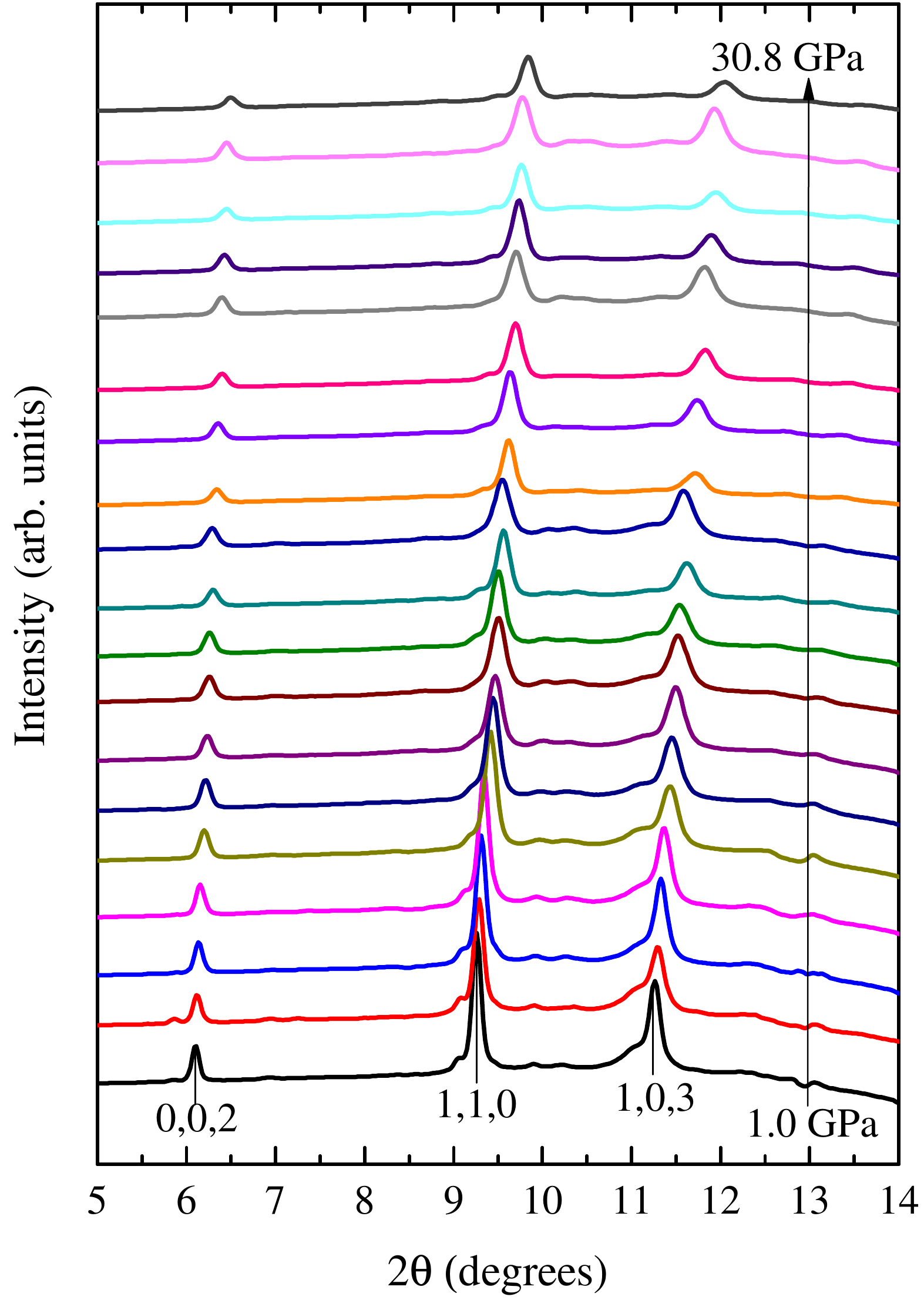}
    \caption{XRD patterns for CePd$_2$P$_2$ at various pressures.  Pressure increases with increasing height, beginning at about \SI{1}{\giga \pascal} and increasing to a maximum of about \SI{30}{\giga \pascal}.  No structural transition is observed within this pressure range.}
 \label{fig:XRD}
\end{figure}

An important question is whether the disappearance of the magnetic transition in $\rho$ and $\chi$ as well as the peak in resistivity as a function of pressure near \SI{12}{GPa} occur within the ambient pressure crystal stucture, as a result of the physics predicted by BKV theory~\cite{belitz_first_1999}, or are instead merely related to a pressure-induced structural transition.
In order to test this, the crystal structure of CePd$_2$P$_2$ was examined via angle-dispersive x-ray diffraction.
X-ray data from an area detector were processed into usable XRD patterns using Dioptas~\cite{prescher_dioptas_2015} 
and then analyzed via GSAS-II~\cite{toby_gsas-ii_2013}.
Figure~\ref{fig:XRD} shows a portion of the results from x-ray diffraction measurements between \SI{1.0}{GPa} and \SI{30.8}{GPa}.
The data show no evidence for any change in the crystal structure to the highest pressures measured.

\begin{figure}
\includegraphics[width=0.9\columnwidth]{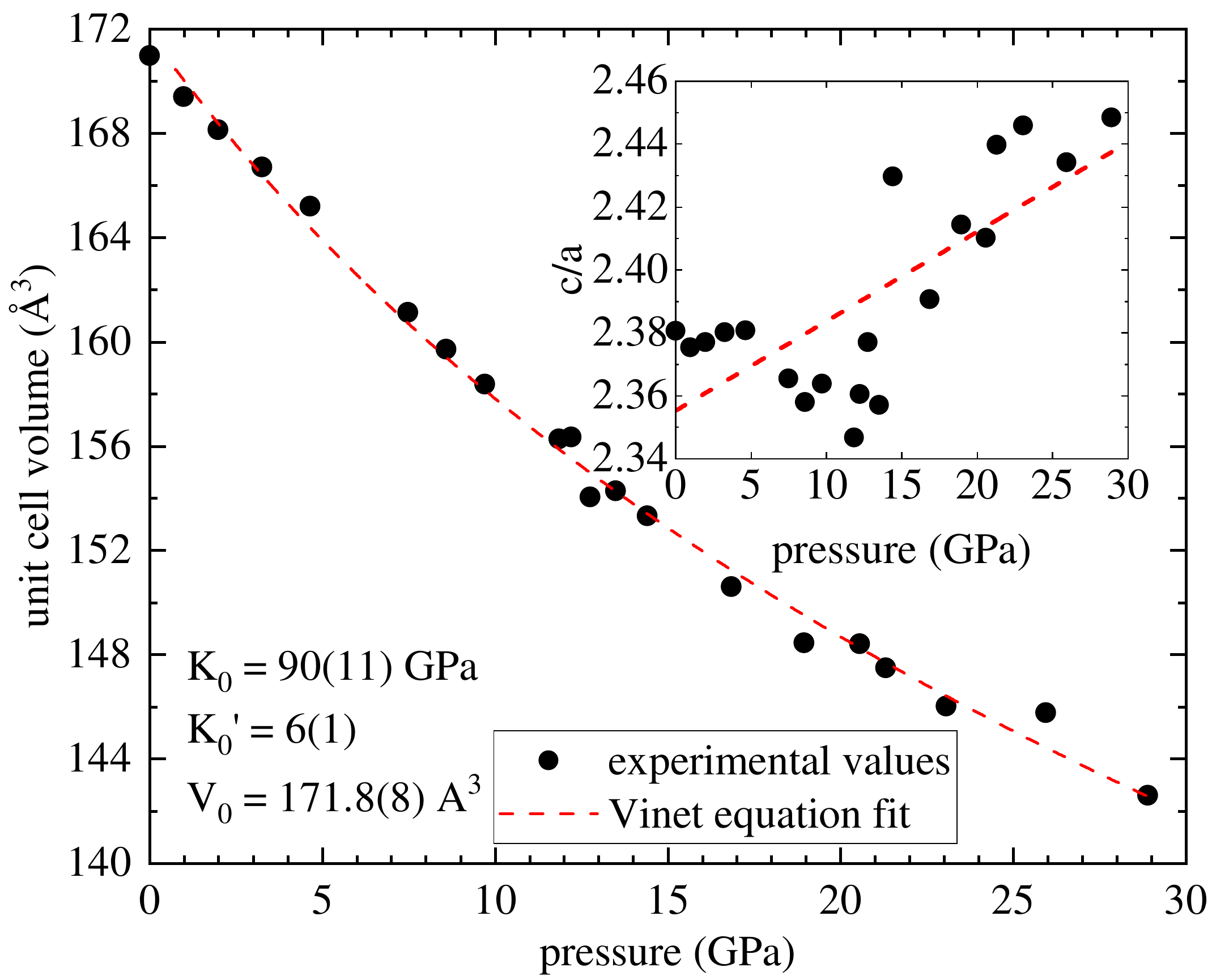}
    \caption{Unit cell volume vs pressure, obtained by x-ray diffraction. The Vinet equation was fitted to the data to obtain the bulk modulus.  The inset shows the ratio of lattice constants $c/a$ as a function of pressure. The $c/a$ ratio increases with pressure, suggesting that CePd$_2$P$_2$ exhibits substantially three-dimensional bonding.  The increased scatter at around \SI{7.4}{\giga \pascal} may be due to the freezing of the pressure medium.  Error bars are approximately the size of the symbols shown.}
 \label{fig:VvsP}
\end{figure}

Data were fit to the ThCr$_2$Si$_2$ structure using LeBail analysis.
Figure~\ref{fig:VvsP}, presents the unit cell volume vs pressure for \ch{CePd2P2}, as determined from these fits.
The Vinet equation~\cite{vinet_temperature_1987} was fit to the equation of state data to obtain a value for the bulk modulus and its derivative with respect to pressure.
Based on the extracted lattice constants a plot of the $c/a$ ratio is shown in the inset.
The $c/a$ ratio increases with increasing pressure, demonstrating that CePd$_2$P$_2$ exhibits a substantial degree of three dimensional bonding.
In a ``layered'' compound, one would expect $c/a$ to decrease under pressure.
The scatter in the $c/a$ ratio increases substantially above \SI{7.4}{GPa}, likely due to the freezing of the pressure medium near this pressure.
There does appear to be a minimum in the $c/a$ ratio near the critical pressure of \SI{12}{GPa}, however it is unclear if this is a consequence of the destruction of the magnetic phase or merely a result of the non-hydrostatic pressure.

\section{Discussion}
\begin{figure}[ht!]
\includegraphics[width=\columnwidth]{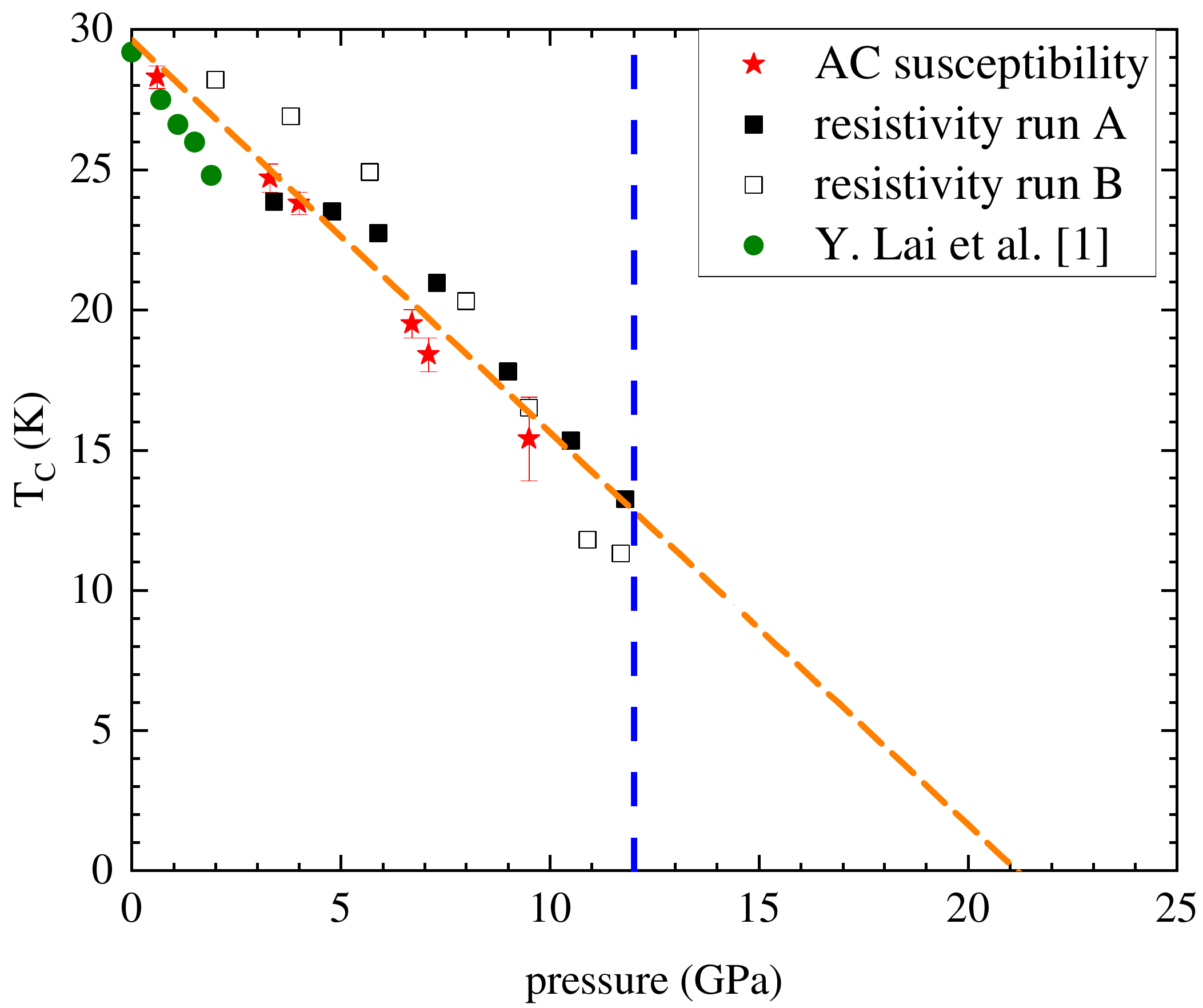}
    \caption{Phase diagram showing the ferromagnetic transition temperature of CePd$_2$P$_2$ vs pressure.  The orange line represents a linear fit to this data, and the vertical blue line indicates the pressure at which the resistivity peaked as a function of pressure.}
 \label{fig:phase_diagram}
\end{figure}

The CePd$_2$P$_2$ phase diagram displayed in Fig.~\ref{fig:phase_diagram} contains $\rm{T_C}$ vs pressure data from Ref.~\cite{lai_ferromagnetic_2018} (up to $\sim \SI{2}{GPa}$) alongside the data described in this work.
The decrease in $\rm{T_C}$ is observed to be nearly linear with increasing pressure.  The scatter in $\rm{T_C}$ is likely due to the quasi-hydrostaticity of the pressure media at high pressure and the different media used in different measurements.
The dashed orange line is a linear fit to all of our data shown in Fig.~\ref{fig:phase_diagram}.
The fit gives a slope of \SI{-1.31(6)}{K/GPa}, a y-intercept of \SI{29.0(3)}{K}, and an x-intercept of \SI{22(1)}{GPa}.
The vertical dashed blue line in Fig.~\ref{fig:phase_diagram} indicates the approximate pressure at which the transition disappears in $\rho$ and $\chi$ and where a peak in the resistivity versus pressure plot was observed.
The general shape of the phase diagram, with a transition that abruptly drops to zero temperature as an order parameter increases (in this case, pressure), is in line with the predictions of BKV theory~\cite{belitz_first_1999}.
However, a central prediction of BKV theory is that this transition is generically of first order, and therefore the second order transition at higher temperature should become first order beyond the tricritical point.
Our data is unable to distinguish whether the transition becomes first order before vanishing.

From the X-ray diffraction data, we can also determine the Ce-Ce nearest neighbor distance; at ambient pressure it sits at approximately \SI{4.2}{\angstrom}, while at the highest measured pressure of \SI{29.8}{\giga \pascal}, it is compressed to \SI{3.9}{\angstrom}.
This is above the cerium Hill limit of \SI{3.6}{\angstrom}.~\cite{moore_nature_2009,miner_plutonium_1971}
On its own, this would suggest that CePd$_2$P$_2$ is a local moment compound, but this is contradicted by the shape of the phase diagram, which, as noted above, fits the predictions of BKV theory, suggesting that CePd$_2$P$_2$ is an itinerant electron compound at high pressure.
The reduction of the signal in the AC susceptibility data at high pressure also suggests that CePd$_2$P$_2$ may be near to an itinerant-to-local transition~\cite{rhodes_effective_1963}.

\begin{figure}[ht!]
\includegraphics[width=\columnwidth]{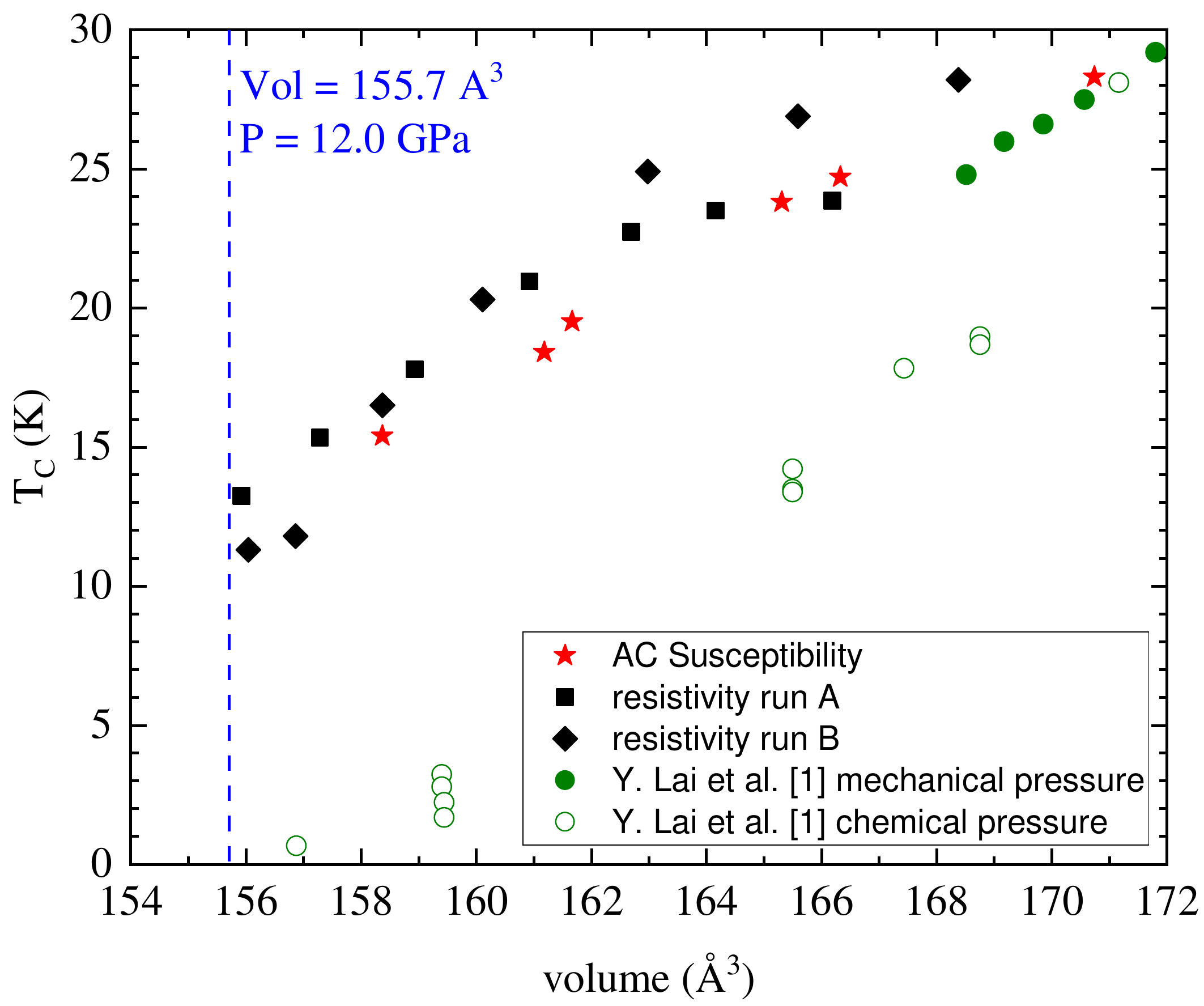}
    \caption{Critical temperature as a function of volume for CePd$_2$P$_2$ and \ch{Ce(Ni,Pd)2P2}.  The filled symbols represent high pressure data on \ch{CePd2P2}, while the open symbols correspond to lattice constants for \ch{Ce(Ni,Pd)2P2} from Ref.~\cite{lai_ferromagnetic_2018}. The magnetic transition is suppressed by chemical and mechanical compression at different rates, though both decrease roughly linearly.  Despite a large difference in $\rm{T_C}$, the magnetic transition becomes undetectable at the same critical volume.}
 \label{fig:VvsVol}
\end{figure}

Figure~\ref{fig:VvsVol} compares the effects on physical pressure on \ch{CePd2P2} and chemical pressure in \ch{Ce(Pd,Ni)2P2} by plotting the ordering temperature vs unit cell volume.
The change in volume from mechanical compression is derived from LeBail analysis of x-ray diffraction data shown in Fig.~\ref{fig:XRD}.
This is compared to unit cell volume data for different levels of Ni substitution reported by Y. Lai \textit{et al.}~\cite{lai_ferromagnetic_2018}.
In both cases, the critical temperature varies roughly linearly with volume, though chemical compression and mechanical compression suppress the transition at different rates.
As the volume decreases, there is a large difference in $\rm{T_C}$ between applied chemical and mechanical pressure.
However, remarkably, in both cases, the transition becomes undetectable at roughly the same critical volume.
It thus appears that \ch{Ce(Ni,Pd)2P2} and \ch{CePd2P2} may offer an ideal pair of systems to compare ferromagnetic quantum phase transitions driven by compression with or without disorder.

There are a number of compounds to which one can compare CePd$_2$P$_2$.
CeTiGe$_3$ is a ferromagnet at ambient pressure, but, as pressure increases, the ferromagnetic transition is suppressed until \SI{4.1}{\giga \pascal}, at which point it becomes (possibly) antiferromagnetic~\cite{kaluarachchi_quantum_2018}.
LaCrGe$_3$ provides yet another example of the possible $T-p-H$ phase diagram that can result from suppressing a ferromagnetic transition.
LaCrGe$_3$ appears to simultaneously exhibit both metamagnetic wings accompanying a shift to a first order transition, as predicted by BKV~\cite{belitz_first_1999}, as well as a modulated magnetic phase ~\cite{kaluarachchi_tricritical_2017,taufour_ferromagnetic_2018}.
Like these compounds, CePd$_2$P$_2$ has a ferromagnetic transition suppressed by pressure until it reaches a critical pressure at which the transition abruptly vanishes.
However, there are no signs of other features such as metamagnetic wings or other magnetic phases in the CePd$_2$P$_2$ phase diagram.
On the other hand, we can not rule out the possibility of a different type of magnetic order in the region between about \SI{9.5}{GPa}, where the anomaly in $\chi$ disappears, and \SI{12}{GPa}, where the resistivity peaks and the anomaly disappears in $\rho$.

\section{Conclusion}
CePd$_2$P$_2$ is a ferromagnet with $\rm{T_C} \approx \SI{29}{K}$ at ambient pressure.
This transition temperature decreases roughly linearly with pressure at a rate of \SI{1.3}{K/GPa}.
The second-order transition is expected to shift to first order and then rapidly vanish, as observed in other clean ferromagnets; the vanishing of $\rm{T_C}$ appears to occur in CePd$_2$P$_2$ near a critical pressure of \SI{12}{GPa}.
This is evidenced by a gradual suppression and then disappearance of the anomaly at $\rm{T_C}$ in magnetic susceptibility and then a broadening and disappearance of the anomaly in the electrical resistivity.
In addition, the apparent disappearance of magnetic order is accompanied by a peak in resistivity and in the magnitude of the magnetoresistance at the critical pressure.
However, the shift to a first order transition could not be directly verified through the present measurements.
High pressure x-ray diffraction shows that these features are not connected to a structural transition and that the ambient pressure crystal structure is maintained to at least \SI{30}{GPa}.
At pressure above the critical pressure we find no clear evidence for metamagnetic wings (as observed in \textit{e.g.}, UGe$_2$~\cite{taufour_tricritical_2010,kotegawa_evolution_2011}, ZrZn$_2$~\cite{uhlarz_quantum_2004,kabeya_non-fermi_2012}, and LaCrGe$_3$~\cite{kaluarachchi_tricritical_2017,taufour_ferromagnetic_2018}), or alternative magnetic structures (as observed in \textit{e.g.}, CeRuPO~\cite{kotegawa_pressuretemperaturemagnetic_2013,lengyel_avoided_2015}, CeTiGe$_3$~\cite{kaluarachchi_quantum_2018} and LaCrGe$_3$~\cite{kaluarachchi_tricritical_2017,taufour_ferromagnetic_2018}).
High pressure neutron scattering measurements could help to definitively determine the microscopic nature of the magnetic order in the vicinity of the critical pressure and whether magnetic order persists in the region beyond the critical pressure.

\section{Acknowledgments}
This work was supported by National Science Foundation (NSF) CAREER award DMR-1453752.  High pressure technique development was partially supported by a National High Magnetic Field Laboratory User Collaboration Grant.  The National High Magnetic Field Laboratory is supported by the NSF via Cooperative agreement No.\ DMR-1157490, the State of Florida, and the U.S. Department of Energy. Designer diamond anvils were supported by DOE-NNSA Grant No.\ DE-NA-0002928 and under the auspices of the U.S. Department of Energy by Lawrence Livermore National Laboratory under Contract DE-AC52-07NA27344.
Portions of this work were performed at HPCAT (Sector 16), Advanced Photon Source (APS), Argonne National Laboratory. HPCAT operations are supported by DOE-NNSA’s Office of Experimental Sciences.  The Advanced Photon Source is a U.S. Department of Energy (DOE) Office of Science User Facility operated for the DOE Office of Science by Argonne National Laboratory under Contract No. DE-AC02-06CH11357.
TAE, DV, and JJH thank Yue Meng for her aid in measurements performed at Argonne National Laboratory.
Dioptas and GSAS-II programs were used for analysis of x-ray diffraction data.
RB and YL acknowledge support from Department of Energy through the Center for Actinide Science and Technology (an EFRC funded under Award DE-SC-0016568).
Designer diamond anvils were supported by DOE-NNSA Grant No. DE-NA-0003916.
\bibliography{CePd2P2}
\end{document}